\documentstyle[12pt]{article}

\title{ \ \\[-7ex] \rule{\textwidth}{1pt} \\ {\bf 
Cluster renormalization in the \BD\ equations} \\[-1ex] }
\author{Peter V Coveney$\dag$ and 
Jonathan AD Wattis$\ddag$ \\[-0.5ex] 
{\footnotesize $\dag$Centre for Computational Science, 
Queen Mary \& Westfield College, Mile End Road, London, E1 4NS.}\\[-0.5ex]
{\footnotesize $\ddag$Division of Theoretical Mechanics, 
School of Mathematical Sciences,}\\[-0.5ex] 
{\footnotesize University of Nottingham, 
University Park, Nottingham, NG7 2RD.}\\[-0.5ex]
{\footnotesize $\dag$ 
\verb$p.v.coveney@qmw.ac.uk$  \hspace*{8mm} 
$\ddag$ \verb$Jonathan.Wattis@nottingham.ac.uk$}\\[-2ex] }
\date{{\footnotesize 14 July 1999}
\\[-1ex] \rule{\textwidth}{1pt}}

\newcommand{\beqa}{\begin{eqnarray}}
\newcommand{\eeqa}{\end{eqnarray}}
\newcommand{\beq}{\begin{equation}}
\newcommand{\eeq}{\end{equation}}
\newcommand{\rec}[1]{\mbox{$\frac{1}{#1}$}}

\newcommand{\half}{\mbox{$\frac{1}{2}$}}

\newcommand{\ds}{\displaystyle}
\newcommand{\BD}{Becker-D\"{o}ring}

\newcommand{\ro}{\varrho}
\let\tilde=\widetilde

\newcommand{\rsum}{\sum_{r=1}^{\infty}}

\newcommand{\ep}{\varepsilon}
\newcommand{\de}{\delta}
\newcommand{\erfc}{{\rm erfc}}

\newcommand{\De}{\Delta}

\newcommand{\lbl}[1]{\label{#1}}

\begin{document}
\maketitle 
\begin{center} {\bf Abstract} \end{center} 
\vspace*{-2mm}

{\small
We apply ideas from renormalization theory to models of cluster 
formation in nucleation and growth processes.   We study a simple 
case of the \BD\ system of equations and show how a novel 
coarse-graining procedure applied to the cluster aggregation 
space affects the coagulation and fragmentation rate coefficients.  
A dynamical renormalization structure is found to underlie the \BD\ 
equations, nine archetypal systems are identified, and their 
behaviour is analysed in detail. These architypal systems divide 
into three distinct groups: coagulation-dominated systems, 
fragmentation-dominated systems and those systems where the two 
processes are balanced. The dynamical behaviour obtained for these 
is found to be in agreement with certain fine-grained solutions 
previously obtained by asymptotic methods. This work opens the way 
for the application of renormalization ideas to a wide range of 
non-equilibrium physicochemical processes, some of which we 
have previously modelled on the basis of the \BD\ equations.}

\subsubsection*{I \ Introduction}

In this paper, we study the \BD\ cluster kinetic equations
familiar from classical nucleation theory \cite{bd35} in which 
the monomer concentration ($c_1$) is held constant
\beq
\dot c_r = J_{r-1} - J_r,\;\; (r\geq2), \qquad
J_r = a_r c_r c_1 - b_{r+1} c_{r+1} . \lbl{fullBD}  
\eeq
Here $c_r$ represents the concentration of clusters 
containing $r$-monomers, the dot implies a time-derivative, 
and $J_r$ is the flux from clusters of size $r$ to those of 
size $r+1$. There are certain mathematical properties of the 
\BD\ system not immediately apparent from the equations but 
crucial to its wide-ranging physical applicability. Firstly, 
the partition function, $Q_r$, satisfies $a_rQ_r = b_{r+1} 
Q_{r+1}$ together with $Q_1=1$ and formally yields the 
equilibrium solution $c_r^{{\rm eq}}=Q_rc_1^r$.  The relevance 
of this solution depends on the behaviour of $Q_r$ 
in the limit $r\rightarrow\infty$; further analysis  
of this is given in Section V, where specific examples 
are analysed in detail.  The function 
\beq
V(\{c_r\}) = \ds\rsum c_r \left( \log \left( 
\ds\frac{c_r}{Q_rc_1^r}\right) - 1 \right) , \lbl{V1eq}
\eeq
is monotonically decreasing and, provided it is bounded 
below, qualifies as a Lyapunov function guaranteeing the 
convergence of arbitrary initial data to the equilibrium 
solution.  We note that the density $\ro=\rsum rc_r$ is not 
constant since monomers can be added to or removed from the system. 
Finally there is a `weak form' for the first equality in (\ref{fullBD})
\beq
\sum_{r=2}^\infty g_r \dot c_r = g_1 J_1 + 
\rsum \,[\, g_{r+1} - g_r \,]\, J_r . \lbl{ids} 
\eeq
In forthcoming analysis we assume initial conditions 
($c_r(0)$) which for large aggregation numbers decay faster 
than any exponential in $r$.

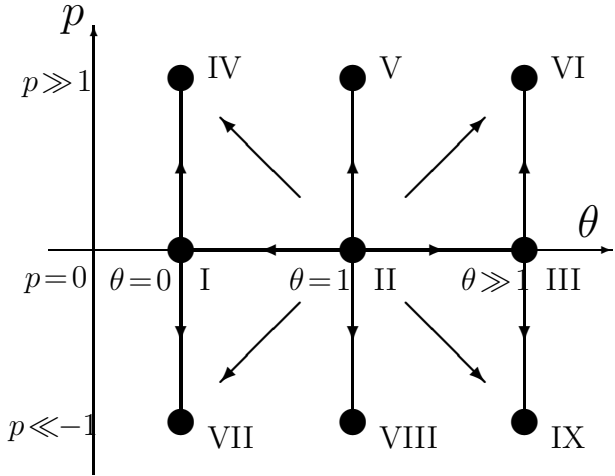
\begin{figure}[bt]
\vspace{2.0in}
\begin{picture}(300,30)(30,20)
\thinlines
\put( 50,100){\vector(1,0){213}}
\put( 67, 15){\vector(0,1){170}}
\put(250,105){{\Large \mbox{$\theta$}}}
\put( 55,185){{\Large \mbox{$p$}}}
\put(100,100){\circle*{10}}
\put(165,100){\circle*{10}}
\put(230,100){\circle*{10}}
\put(100, 35){\circle*{10}}
\put(165, 35){\circle*{10}}
\put(230, 35){\circle*{10}}
\put(100,165){\circle*{10}}
\put(165,165){\circle*{10}}
\put(230,165){\circle*{10}}
\thicklines
\put(165,100){\vector( 1, 0){35}}
\put(165,100){\vector( 0, 1){35}}
\put(165,100){\vector(-1, 0){35}}
\put(165,100){\vector( 0,-1){35}}
\put(200,100){\line  ( 1, 0){30}}
\put(130,100){\line  (-1, 0){30}}
\put(165,135){\line  ( 0, 1){30}}
\put(165, 65){\line  ( 0,-1){30}}
\put(100,100){\vector(0, 1){35}}
\put(100,100){\vector(0,-1){35}}
\put(100,135){\line  (0, 1){30}}
\put(100, 65){\line  (0,-1){30}}
\put(230,100){\vector(0, 1){35}}
\put(230,100){\vector(0,-1){35}}
\put(230,135){\line  (0, 1){30}}
\put(230, 65){\line  (0,-1){30}}
\put(185,120){\vector( 1, 1){30}}
\put(145,120){\vector(-1, 1){30}}
\put(185, 80){\vector( 1,-1){30}}
\put(145, 80){\vector(-1,-1){30}}
\put(107, 85){I}
\put(173, 85){II}
\put(238, 85){III}
\put(110,165){IV}
\put(175,165){V}
\put(240,165){VI}
\put(110, 25){VII}
\put(175, 25){VIII}
\put(240, 25){IX}
\put( 73, 85){$\theta\!=\!0$}
\put(141, 85){$\theta\!=\!1$}
\put(206, 85){$\theta\!\gg\!1$}
\put( 41, 86){$p\!=\!0$}
\put( 40,160){$p\!\gg\!1$}
\put( 35, 30){$p\!\ll\!\!-1$}
\end{picture}
\caption{{\small The effect of our coarse-graining  
dynamical renormalization on the two parameters 
$\theta,p$ in a \BD\ model with rate coefficients which 
vary as a power $p$ of the cluster size $r$, and with 
ratio of aggregation to fragmentation rates $\theta$. 
The dots show the fixed points of the mapping.}}
\label{rggraph}
\end{figure}

\subsubsection*{II \ Coarse-graining of cluster aggregation space}

We now perform a coarse-graining contraction of the infinite 
set of \BD\ equations by systematically eliminating all the 
concentration variables except those which represent an 
aggregation number $\Lambda_r$ where 
\beq
\Lambda_r=(r-1)\lambda+1, \;\;\;\;\;r=1,2,3,\ldots.
\lbl{mesh} \eeq
We then relabel the retained concentrations by 
$x_r = c_{\Lambda_r}$.  The reduced fluxes are  
\beqa 
L_r & = & \alpha_r x_r x_1^{\lambda_r} - 
\beta_{r+1} x_{r+1} , \lbl{cgBDL} \\
\alpha_r & = & 
T a_{\Lambda_r}a_{\Lambda_r+1}\ldots a_{\Lambda_{r+1}-1},
\lbl{cgalpha} \\ \beta_{r+1} & = &  
T b_{\Lambda_r+1} b_{\Lambda_r+2} \ldots b_{\Lambda_{r+1}},
\lbl{cgbeta} \eeqa
where $T$ is a constant which represents a change 
of timescale; the kinetic equations then reduce to 
\beq
\dot x_r = L_{r-1} - L_r ,\;\;\; (r\geq2). 
\lbl{cgBD} \eeq
This procedure is analogous to the Kadanoff block-spin 
renormalization procedure \cite{goldenfeld}; detailed 
information for cluster sizes between the aggregation 
numbers $\Lambda_r$ is lost.
For more details of this procedure, see \cite{cw96,wk98}.  
If the contracted system is to faithfully approximate 
the original system, we require that the special 
mathematical properties mentioned above are preserved under 
the coarse-grained rescaling. We can then draw on our  
renormalization procedure to extract the structurally 
stable phenomena present in the system.

The physical properties of the full \BD\ system 
(\ref{fullBD}) are shared by the contracted system 
(\ref{cgBDL})--(\ref{cgBD}):  the partition function satisfies 
$\alpha_r Q_{\Lambda_r} = \beta_{r+1} Q_{\Lambda_{r+1}}$,  
hence $x_r^{{\rm eq}} = Q_{\Lambda_r} x_1^{\Lambda_r}$ is 
formally an equilibrium solution. The function $V(\{x_r\})=\rsum 
x_r(\log(x_r/Q_{\Lambda_r}x_1^{\Lambda_r})-1)$ has the same 
properties as (\ref{V1eq}).  The weak form (\ref{ids}) is 
still valid if $c_r$ is replaced by $x_r$ and $J_r$ by $L_r$. 
Finally the density in the system is now defined by 
\beq
\ro = x_1 + \lambda \ds \rsum [(r-1)\lambda+1] x_r .
\lbl{conrodef} \eeq
To apply renormalization ideas to this theory, we consider the 
repeated application of the coarse-graining transformation 
(\ref{cgBDL})--(\ref{cgBD}), so we now reapply the contraction 
procedure with mesh size $\mu$.  Defining new variables $z_1=x_1$,  
$z_r=x_{(r-1)\mu+1}$, and $I_r$ as the flux from $z_r$ to $z_{r+1}$, 
we find
\beq 
\dot z_r = I_{r-1} - I_r, \;\;\;(r\geq2), \hspace*{5mm}
I_r = A_r z_r z_1^{\lambda\mu} - B_{r+1} z_{r+1} , 
\eeq
with $A_r,B_r$ determined from $\alpha_r,\beta_r$ 
in an analogous way to (\ref{cgalpha})-(\ref{cgbeta}). 
A similar set of physical properties holds for this 
system of equations as for the original \BD\ equations. 
Thus {\em a repetition of the coarse-grained contraction is 
identical to a single application with a larger mesh 
parameter} $\lambda\mu$. This shows that it is sufficient 
to consider a system of equations which has undergone 
a single contraction with large $\lambda$.

\subsubsection*{III \ The case of constant coefficients}

Although ultimately a theory capable of handling arbitrary 
forms of rate coefficients $a_r$ and $b_{r+1}$ is our goal, 
for the sake of simplicity let us start by considering 
constant coefficients -- that is $a_r=a$, $b_r=b$. The 
parameter $\theta=ac_1/b$ enables the system's behaviour 
to be classified.   The cluster partition function is defined by 
$Q_r=(a/b)^{r-1}$ and the forward coefficients in the 
reduced model by $\alpha_r = T a^\lambda$, $\beta_{r+1} = 
T b^\lambda$.  Thus the size-independent rate coefficients 
$a_r=a$, $b_r=b$ are mapped to size-independent rate 
coefficients in the reduced model.  This coarse-graining 
maps $\theta$ to $\theta^\lambda$, leading to three fixed 
points, $\theta=0,1,\infty$.  The large-time asymptotics 
of systems with constant coefficients have been analysed 
in detail in \cite{wk98}, where it is shown that the 
$\theta=0$ case converges to the equilibrium solution, the 
$\theta=\infty$ case converges to the steady-state solution 
$x_r=x_1$ by a diffusive wave which moves through aggregation 
space in such a way that its position is given by 
$r=s(t)\sim t$, and the $\theta=1$ case converges to the 
equilibrium solution $x_r=x_1$ by purely diffusive means 
($x_r\sim x_1 \erfc(r/2\sqrt{t}$)).

Ref \cite{wk98} also shows that in order for the contracted 
system to preserve the correct large-time asymptotics, the 
parameter $T$ in (\ref{cgalpha})-(\ref{cgbeta}) should take 
the value 
\beq
T=\frac{ac_1-b}{\lambda(a^\lambda c_1^\lambda-b^\lambda)};
\lbl{Tresc} \eeq
this temporal rescaling implies that our renormalization is 
{\em dynamic} \cite{goldenfeld}. Following this temporal rescaling, 
the large-time limit of the density of the original system with 
$\theta>1$ then scales with $\half c_1 (ac_1\!-\!b) t^2$, which is 
identical to the result given by the coarse-grained system 
(\ref{conrodef}); and both $V(\{c_r\})$ of equation (\ref{V1eq}) 
and $\lambda V(x)$ scale with $-\half c_1 (ac_1\!-\!b)^2 t^2 \log\theta$.

\subsubsection*{IV \ The case of power-law rate coefficients}

In many systems the reaction rates are not independent of 
size as assumed above, but rather depend on the size of 
the cluster according to some power law. We assume $a_r = 
ar^p$, $b_{r+1}=br^p$, allowing us to model surface-limited 
aggregation in $d$ dimensions with $p=1-1/d$. The 
parameter $\theta=ac_1/b$ remains a useful tool for classifying 
behaviour; the partition function remains $Q_r = (a/b)^{r-1}$. 
The forward coefficients in the reduced model are 
\beq
\alpha_r = a^\lambda \left\{ [(r-1)\lambda+1] 
[(r-1)\lambda+2] \ldots [r\lambda]  \right\}^p . 
\eeq
For asymptotically large $\lambda$ these can be approximated by 
\beq
\log \alpha_r \sim \lambda\log a + p \lambda 
\,\left[\, \log(r\lambda) - 1 + (1\!-\!r) \log \left( 
1 - \rec{r} \right) \right] , \lbl{approx} 
\eeq
so for simplicity we shall take $\alpha_r=(a\lambda^pr^p)^\lambda$, 
which is asymptotically correct at large $r$ and differs only 
slightly at small values of $r$.  In the same manner, the backward 
rate coefficients in the contracted model are given by $\beta_{r+1} = 
(b\lambda^p r^p)^\lambda$.

Our coarse-graining contraction maps the set of 
models with power law rate coefficients into itself.  
The coarse-graining of power-law coefficients is 
only approximate when $p\neq0$. However, the large-time 
asymptotics is qualitatively preserved, provided that a 
similar temporal rescaling is performed as in equation 
(\ref{Tresc}) \cite{rgfull}. 
For any given model the contraction maps the exponent 
$p$ to $\lambda p$.  Following a contraction with 
large $\lambda$, there are three cases to consider: 
$p=0$, and large positive or negative $p$.  The 
reduced system also has a different $\theta$-parameter, 
$\tilde{\theta} = \alpha_r x_1^\lambda/\beta_{r+1} = 
\theta^\lambda$; thus the contraction maps $\theta$ to 
$\theta^\lambda$. The fixed points $\theta=0,1,\infty$ 
are therefore of most interest to us.  Combining this 
information, there are nine fixed points of the 
coarse-grained contraction in $(\theta,p)$ parameter space, 
and these form the basis of the ensuing analysis.

Figure \ref{rggraph} shows schematically the effect 
of the contraction.   In phase plane terminology, 
II has the form of an unstable node, I, III, V, VIII are 
saddle points (although they are at the limits of the 
allowable domain, so only have trajectories on one 
side of the fixed point), and IV, VI, VII, IX are 
stable nodes.   Cases I, IV, VII all have partition 
function $Q_r=0$ for $r\geq2$; in cases II, V, VIII
the partition function satisfies $Q_r=1$, whilst 
it is undefined in cases III, VI, IX since in all 
these cases the fragmentation rate is zero.  
Having no equilibrium configuration,  these 
three cases approach a steady-state solution.

\subsubsection*{V \ Effect of perturbations on the fixed points}

There are two reasons for wanting to study noisy coefficients: 
firstly any set of coefficients will be subject to uncertainties, 
whether derived from experimental data or a mathematical model.  
Secondly, systems are always susceptible to thermal (and in the 
models we study also spatial) fluctuations which locally alter 
the rate coefficients. In both cases it is necessary to know whether 
the models used are stable to minor variations in rate coefficients.

Firstly we allow each reaction rate ($a_r,b_{r+1}$ for $r=1,2,\ldots$) 
to be independently perturbed by a small amplitude random fluctuation 
of the form 
\beq
a_r = a r^p ( 1 + \nu \xi_r ) , \hspace*{5mm} 
b_{r+1} = b r^p ( 1 + \nu \chi_{r+1} ) , \qquad r=1,2,\ldots . 
\eeq 
with $\nu\ll1$ and $\xi_r,\chi_{r+1}$ being independent random 
variables with zero mean satisfying $\xi_r,\chi_{r+1}={\cal O}(1)$. 
Such perturbations have no effect on the leading order equilibrium 
or steady-state solutions, or the large time asymptotics.

A more interesting case is that in which the presence of noise 
in the rate coefficients is allowed to alter their leading order 
behaviour at large $r$. To examine these, we perturb the forward 
and backward rate coefficients according to 
\beq 
a_r = a r^p + \de_r , \hspace*{5mm} 
b_{r+1} = b r^p + \ep_{r+1} , \qquad r=1,2,\ldots . 
\lbl{noise} \eeq 
where $\de_r,\ep_{r+1}$ have characteristic magnitude $\nu\ll1$. 
We now investigate the effect of such perturbations on the 
equilibrium and steady-state solutions.  Since we assume 
$c_1=1$, the partition function is the equilibrium solution. 
However, as described in \cite{wk98}, there are cases where the 
equilibrium configuration formally has infinite mass and is hence 
not relevant;  the system then approaches one of the family of 
steady-state solutions in which all fluxes, $J_r$, are equal. 
The steady-state flux is determined by requiring the most 
rapid decay in $c_r$ as $r\rightarrow\infty$.  We now apply 
these ideas to the nine fixed points isolated earlier.

{\em Case I: $p=0$, $\theta=0$}. 
Since the non-perturbed case has $a_r=0$ for all $r$, 
the partition function and the equilibrium solution are 
then zero; introducing perturbations removes this 
degeneracy, and the equilibrium solution then 
becomes rapidly decaying in $r$, namely 
$c_r={\cal O}(\nu^{r-1})$, where $\nu$ is 
a small parameter representing the typical 
size of perturbations $\delta_r$.

{\em Case II: $p=0$, $\theta=1$}. 
In this case,  introducing perturbations to the rates 
modifies the partition function from $Q_r=1$ to 
\beq Q_r \sim 1 + \sum_{k=1}^r (\de_k-\ep_{k+1}) , \eeq
so small amplitude noise in the coefficients does not affect 
the leading-order behaviour of the system.

{\em Case III: $p=0$, $\theta=\infty$}. 
In the absence of perturbations, there is no partition function 
for this case; when present, $Q_r \sim \prod_{k=1}^{r-1} 
(1/\ep_k)$. However, this case converges to a steady-state
rather than the equilibrium. When perturbations are included, 
the steady flux is $J=1+(\de_1\!-\!\ep_2)+{\cal O}(\nu^2)$, 
which implies that the concentrations asymptote to 
$c_r = 1 + (\de_1\!-\!\ep_2\!+\!\ep_{r+1}\!-\!\de_r) + 
{\cal O}(\nu^2)$.

{\em Case IV: $p\gg1$, $\theta=0$}. 
As in case I, where noise is absent the partition 
function, $Q_r$,  is zero for $r\geq2$. Introducing 
noise removes this degeneracy, for small $\nu$, 
$Q_r \sim {\cal O}(\nu^{r-1})$.  Thus, as in Case I, 
the equilibrium solution rapidly decays with $r$.

{\em Case V: $p\gg1$, $\theta=1$}. 
The balance of aggregation and fragmentation 
implies that $Q_r\equiv1$ in the case with no noise. 
The addition of noise to the rates alters this, to
\beq
Q_r = 1 + \sum_{k=1}^{r-1} \left( 
\frac{\de_k-\ep_{k+1}}{k^p} \right) + {\cal O}(\nu^2) , 
\eeq
wherein we see that the alteration to the partition function 
only affects the ${\cal O}(\nu)$ correction term, leaving 
the leading order behaviour ($Q_r\sim1$) unaltered.

Note that if $p>1$ then the system does not evolve to the 
equilibrium solution, but instead is attracted to a steady-state 
solution with more rapid decay in the limit $r\rightarrow\infty$. 
Perturbing the rate coefficients modifies this state to 
\beq
c_r = 1 - \frac{1}{\zeta(p)} \sum_{k=1}^{r-1} \frac{1}{k^p} 
+ \sum_{k=1}^{r-1} \frac{\de_k - \ep_{k+1} - J_1}{k^p} , 
\eeq
where $J_1 = (1/\zeta(p)) \sum_{k=1}^\infty (\de_k\!-\!\ep_{k+1})
/k^p$, which has constant flux $J=1/\zeta(p) + J_1$.

{\em Case VI: $p\gg1$, $\theta=\infty$}. 
In this case the system approaches a steady-state 
solution,  with flux $J=1 + (\de_1-2^{-p}\ep_2) + 
{\cal O}(\nu^2)$, implying 
\beq
c_r = \frac{1}{r^p} \left[ 1 + \left( \de_1 - 2^{-p} \ep_2 +
\frac{\ep_{r+1}}{(r\!+\!1)^p} - \frac{\de_r}{r^p} \right) 
\right] . \eeq
Thus noise in the rate coefficients has a minor effect 
on the solution.

{\em Case VII: $p\ll-1$, $\theta=0$}. 
Formally we have, 
\beq
Q_r = \prod_{k=1}^{r-1} \frac{\de_k}{k^p + \ep_{k+1}} , 
\lbl{8mag} \eeq
thus when $r={\cal O}(1)$, $Q_r={\cal O}(\nu^{r-1})$. 
However, when $r=r_c := {\cal O}(\nu^{1/p})$ 
\beq
c_r \sim [(\nu^{1/p})!]^{-p} \exp ( \nu^{1/p} \log \nu ) . 
\eeq
For $r\geq r_c$, the perturbations have the same 
magnitude as the non-random part of the rate 
coefficient, thus all subsequent $Q_r$ values depend 
strongly on the perturbations $\de_k,\ep_k$ and 
have the order of magnitude given by (\ref{8mag}).

{\em Case VIII: $p\ll-1$, $\theta=1$}. 
In the noiseless case this system converges to the equilibrium 
solution $c_r=1$. When noisy coefficients are introduced, this 
solution may cease to be valid since at large $r$, the noise 
will be a leading-order effect.  For small $r$ we construct an 
asymptotic approximation to the modified equilibrium solution
\beq
c_r = 1 + \sum_{k=1}^{r-1} k^{-p} (\de_k\!-\!\ep_{k+1}) + 
{\cal O}(\nu^2) . 
\eeq
This approximation to the solution ceases to be 
valid at large $r$, where $c_r={\cal O}(1)$. We expect 
$c_r$ to remain ${\cal O}(1)$ for all values of $r$, but to 
vary from $c_r=1$ by significant amounts at large $r$.

{\em Case IX: $p\ll-1$, $\theta=\infty$}. 
In the absence of noise, this case approaches the divergent 
steady-state $c_r = r^{-p}$ (with flux $J=1$).  For small 
amplitude noise, a modified form of this solution persists 
\beq
c_r\!=\! \frac{1}{r^p} \left[ 1 + \left( \de_1\!-\!2^{-p}\ep_2
\!-\!\frac{\de_r}{r^p}\!+\!\frac{\ep_{r+1}}{(r\!+\!1)^p}
\right)\right]+{\cal O}(\nu^2) ; 
\eeq
however, this ceases to be valid when 
$r={\cal O}(\nu^{1/p})$. For values of $r$ of this magnitude 
and larger, perturbations to the rates cannot be neglected as
they constitute a leading-order effect in the system; and 
$c_r = {\cal O}(1/\nu)$ for all $r\geq {\cal O}(\nu^{1/p})$.

\subsubsection*{VI Effect of perturbations on the 
coarse-grained reaction rates}

In this section we examine the effect which the coarse-graining 
contraction procedure has on the perturbed rate coefficients. In
particular,  
we investigate whether small amplitude noise in the full 
description of the model maps to small amplitude noise 
in the reduced description.   On inserting (\ref{noise}) into 
(\ref{cgalpha})-(\ref{cgbeta}) with $T=\lambda^{-p\lambda}$, 
we obtain 
\beq
\alpha_r = a^\lambda r^{p\lambda} + \De_r , \hspace*{5mm} 
\beta_{r+1} = b^\lambda r^{p\lambda} + E_{r+1} , 
\eeq
where $\De_r,E_{r+1}$ represent the perturbations in the 
contracted descriptions and depend respectively on the 
$\de_k,\ep_k$.  For each of the nine fixed points (in which 
$a,b=0,1$) we calculate the leading order form of this dependence.

{\em Case I: $p=0$, $\theta=0$}. 
Since $a=0$, $b=1$, we have $\alpha_r=\De_r= 
{\cal O}(\nu^\lambda)$, and $\beta_{r+1}=1+{\cal O}(\nu)$ 
Thus following contraction, the perturbations remain small.

{\em Case II: $p=0$, $\theta=1$}. 
Following the coarse-graining contraction, the reaction 
rates are given by $\alpha_r,\beta_r = 1 + {\cal O}(\nu)$. 
So the perturbations remain the same order of magnitude 
in the contracted model as in the full.

{\em Case III: $p=0$, $\theta=\infty$}. 
The domination of aggregation is not altered by the 
presence of small noise, since $\alpha_r=1+{\cal O}(\nu)$ 
and $\beta_r={\cal O}(\nu)$.

{\em Case IV: $p\gg1$, $\theta=0$}. 
The contracted rates are given by 
$\alpha_r=r^{p\lambda}+{\cal O}(\nu)$ and  
$\beta_{r+1}={\cal O}(\nu)$; 
in the latter, we have made the approximation 
(\ref{approx}) valid for large $r$. 
The system remains fragmentation-dominated.

{\em Case V: $p\gg1$, $\theta=1$}. 
For large $r$, the rates in the contracted system have 
the form $\alpha_r,\beta_{r+1}=r^{p\lambda}+{\cal O}(\nu)$
Thus the noise will not cause any change to the leading 
order form of the rate coefficients.

{\em Case VI: $p\gg1$, $\theta=\infty$}. 
The domination of aggregation persists, since following 
contraction $\alpha_r = r^{p\lambda} + {\cal O}(\nu)$ 
whilst $\beta_{r+1}={\cal O}(\nu^\lambda)$.

{\em Cases VII--IX: $p\ll-1$}. 
The formulas for $\De_r,E_{r+1}$ in these cases are 
identical to cases IV,V,VI respectively. However, here 
$p<0$ so that at large cluster sizes $r$, the perturbations 
will be of the same order of magnitude as the deterministic 
part of the rate coefficients.  This occurs when 
$r={\cal O}(\nu^{1/p})$.

In Cases I--VI, the noise indeed remains small in the 
contracted description of the model hence these may be 
termed universality classes, whilst in Cases VII-IX, 
this is not the case.   In these last three cases, at 
large aggregation numbers, the noise in the full 
description is not small relative to the power law 
component of the rate coefficient, and this is reflected 
in the contracted model.  In Cases VII--IX perturbations 
to the power-law rate coefficients play a major role in 
the kinetics at large cluster sizes $r$, as they do in 
the full model.  Thus Cases VII-IX may be termed 
universality classes if the added noise decays faster 
than the given power law as $r\rightarrow\infty$.

\subsubsection*{VII \ Conclusions}

We have applied renormalization ideas to the \BD\ model of 
cluster-formation.  A novel feature of this work is that it is 
the cluster aggregation space which is rescaled, rather than a 
spatial dimension. Moreover, a dynamical renormalization 
is required to correctly maintain the time-scales of the growth 
and fragmentation processes following the rescaling of aggregation 
space.  In the case of the power-law model, nine fixed points 
of the renormalization procedure have been identified and 
analysed in greater detail, quantitatively providing nine types 
of large-time asymptotics which may be exhibited by the system.  
Five of these systems tend to equilibrium, and the remaining 
four to steady-state solutions.   The pure fragmentation cases 
(I,IV,VII) all tend to the trivial equilibrium $x_r=\delta_{r,1}$.

In Cases I, IV, VII, a diffusive wavefront invades the 
large-$r$ region where cluster concentrations are zero, 
leaving the equilibrium solution behind the wavefront. 
In Cases II, V and VIII the equilibrium solution is approached
by purely diffusive mechanisms, no advection being 
present in the system.  If $p>1$ in Case V, then the system 
approaches a steady-state solution rather than the 
equilibrium solution, since the steady-state has faster 
decay at large aggregation numbers.  This case is thus 
similar to Cases III, VI, IX, all of which approach 
steady-states rather than true thermodynamic equilibrium.  
However, their large-time asymptotics are more akin to Cases 
I,IV,VII, being dominated by a diffusive wave which moves 
into the large $r$-domain.

Thus for the first time we have identified universality 
classes present in the \BD\ equations, in that any system 
with power-law coefficients can be classified into one of 
the nine cases which correspond to fixed points of our 
contraction, and this qualitatively determines the system's 
large-time behaviour. 
In physical terms, our demonstration that a renormalization 
structure underpins the \BD\ equations carries with it the 
implication that universal behaviour can be identified in the 
approach of such systems to equilibrium or steady-states. 
In a forthcoming paper \cite{rgfull}, we shall discuss the 
temporal behaviour in detail.  In the case of the \BD\ equations, 
this is a very welcome development, since it dispenses with 
the need to specify in full detail all the generally unknown 
fine-grained rate coefficients.  For example the partition 
function is left unchanged by the coarse-graining, as is 
the equilibrium solution and the steady-state solution. 
At the end of Section III we showed that the large-time behaviour 
of both the density and the Lyapunov function (free energy) were 
left invariant by our coarse-grain rescaling. 
It is perhaps worth pointing out here, however, that 
consideration of the asymptotic limit implied by the 
renormalization procedure is not necessarily always 
appropriate, e.g. for systems in which it is crucial 
to retain some level of fine-grained detail in order 
to properly capture the dynamics.

The successful application of the renormalization  
techniques reported here opens the way for a study of 
generalised \BD\ equations using similar methods; it 
also furnishes a firm theoretical foundation for the 
analyses we have previously given of various generalisations 
of the basic \BD\ theory to a wide range of processes of 
physicochemical interest, including micelle and vesicle 
formation and self-reproduction \cite{cw96,cw98}, generalised 
nucleation and growth phenomena \cite{wc97}, and 
macromolecular sequence selection in biopolymers \cite{wc98}.

\noindent{\bf Acknowledgements.}
We are indebted to John Cardy for several valuable discussions 
regarding renormalization theory, and to Wolfson College, 
University of Oxford, for hosting us during the development of 
this work. PVC is grateful to Wolfson College and the Department 
of Theoretical Physics, University of Oxford for a Wolfson-Harwell 
Visiting Fellowship (1996-1999).  JADW also wishes to thank 
John King for many very helpful conversations.

\vspace*{-3mm}
\footnotesize

\normalsize

\listoffigures
\end{document}